\documentclass[aps,pre,twocolumn,groupedaddress,showpacs]{revtex4}
\usepackage{graphicx}
\usepackage{amssymb}

\begin{document}

%\preprint{}

\title{Dynamics of Gas-Fluidized Granular Rods}

% address or URL should go in the {}'s for \email and \homepage.

\author{L.J. Daniels$^1$, Y. Park $^2$, T.C. Lubensky$^1$, and D.J. Durian$^1$}
%\thanks{}
\affiliation{$^{1}$Department of Physics and Astronomy, University
of Pennsylvania, Philadelphia, PA 19104-6396, USA}
\affiliation{$^{2}$Department of Physics, Myong Ji University,
Yongin, Kyonggi, Korea}

%Collaboration name if desired (requires use of superscript address

\date{\today}

\begin{abstract}

We study a quasi-two-dimensional monolayer of granular rods
fluidized by a spatially and temporally homogeneous upflow of air.
By tracking the position and orientation of the particles, we
characterize the dynamics of the system with sufficient resolution
to observe ballistic motion at the shortest time scales. Particle
anisotropy gives rise to dynamical anisotropy and superdiffusive
dynamics parallel to the rod's long axis, causing the parallel and
perpendicular mean squared displacements to become diffusive on
different timescales. The distributions of free times and free
paths between collisions deviate from exponential behavior,
underscoring the non-thermal character of the particle motion. The
dynamics show evidence of rotational-translational coupling
similar to that of an anisotropic Brownian particle. We model
rotational-translation coupling in the single-particle dynamics
with a modified Langevin model using non-thermal noise sources.
This suggests a phenomenological approach to thinking about
collections of self-propelling particles in terms of enhanced
memory effects.

\end{abstract}

% list of pacs
%

%\keywords{}

\maketitle

%\section{}
%\subsection{}
%\subsubsection{}

% If in two-column mode, this environment will change to single-column
% format so that long equations can be displayed. Use
%\begin{widetext}
%\end{widetext}

%===============================================================\
\section{Introduction}
Flocking birds, schooling fish, and swarming bacteria are examples
of collections of self-propelled particles -- particles that take
in energy from their environment and then dissipate it by moving
in a preferential direction through that environment -- that
display collective, coherent behavior over a huge range of length
scales. A wealth of theoretical work, ranging from minimal
rule-based models~\cite{Vicsek1995,Chate2004} to coarse-grained
hydrodynamic theories that employ symmetry
considerations~\cite{Tu1995,Tu1998,Ramaswamy2005}, has been
conducted in order to elucidate those behaviors that are universal
to collections of self-propelled particles. The earliest physical
model~\cite{Vicsek1995} predicted a phase transition from a
disordered to a true long-range-ordered state in two dimensions,
in direct contrast to thermal systems where such a transition is
explicitly forbidden. For apolar particles with nematic order, the
broken-symmetry phase is characterized by anomalously large number
fluctuations~\cite{Toner2003,Montagne2006}. Other modes of
collective behavior predicted include swirling and vortex
motion~\cite{Tsimring2003,Tsimring2007} and propagating
waves~\cite{Ulm1998,Ramaswamy2002,Marchetti2007}.

Despite the abundance of models and simulations, little experiment
on physical systems has been conducted, with what has been done
focusing on vertically-vibrated granular
systems~\cite{Kudrolli2003,Nossal2006,Menon2006,Tsimring2007,Menon2007,TsimringPolar}.
These systems afford an advantage over biological systems: there
is control over the microscopics and energy input, and a steady
state can typically be reached provided the system does not age.
Vibrated-bed experiments have observed swirling
vortices~\cite{Kudrolli2003,Tsimring2007}, pattern
formation~\cite{Nossal2006}, and shape-dependent long-range
ordering~\cite{Menon2006}. Recently, giant number fluctuations
were observed in a vertically-vibrated monolayer of
nematic-ordered granular rods~\cite{Menon2007}. Large number
fluctuations were also reported for a collection of vibrated
spheres~\cite{UrbachComment}, fueling some
debate~\cite{MenonResponse} as to whether the large density
fluctuations are due to the self-propelled nature of the
particles, inelastic collisions, or a heaping
instability~\cite{Kudrolli2003} observed in vibrated media.
Additionally, vertical vibration as a method of driving has
several inherent disadvantages: the driving force is temporally
inhomogeneous and the particle dynamics may vary depending on the
phase of the oscillation cycle at the time the particle contacts
the substrate. Further, for smooth plates, a lateral force is only
generated by particle overlap out of plane; particles `boosted' to
high speeds in this manner contribute to power law tails in
velocity distributions.

In the interest of discerning universal characteristics of
collections of self-propelled particles, we consider another
method of driving. Here, we report on a quasi-two-dimensional
monolayer of granular rods fluidized by an upflow of air. The
fluidizing airflow is temporally and spatially homogeneous, with
airspeed less than free-fall and with Reynolds number much greater
than 1. Because of their extended shape, the particles move by
scooting when one of their ends has lifted off the substrate. This
results in self-propelled behavior characterized by dynamical
anisotropy between translations parallel and perpendicular to the
particle's long axis. To elucidate the effect of self-propulsion,
we characterize the dynamics of a single particle and compare it
to the behavior of an anisotropic Brownian
particle~\cite{Perrin1934,Perrin1936,Yodh2006}. To that end, we
model our system using the same Langevin formalism as for a
Brownian particle with a single modification: particle
self-propulsion is included implicitly via non-thermal noise
terms.

\section{Experimental Details}\label{experiment}
\begin{figure*}
\includegraphics[width=7in]{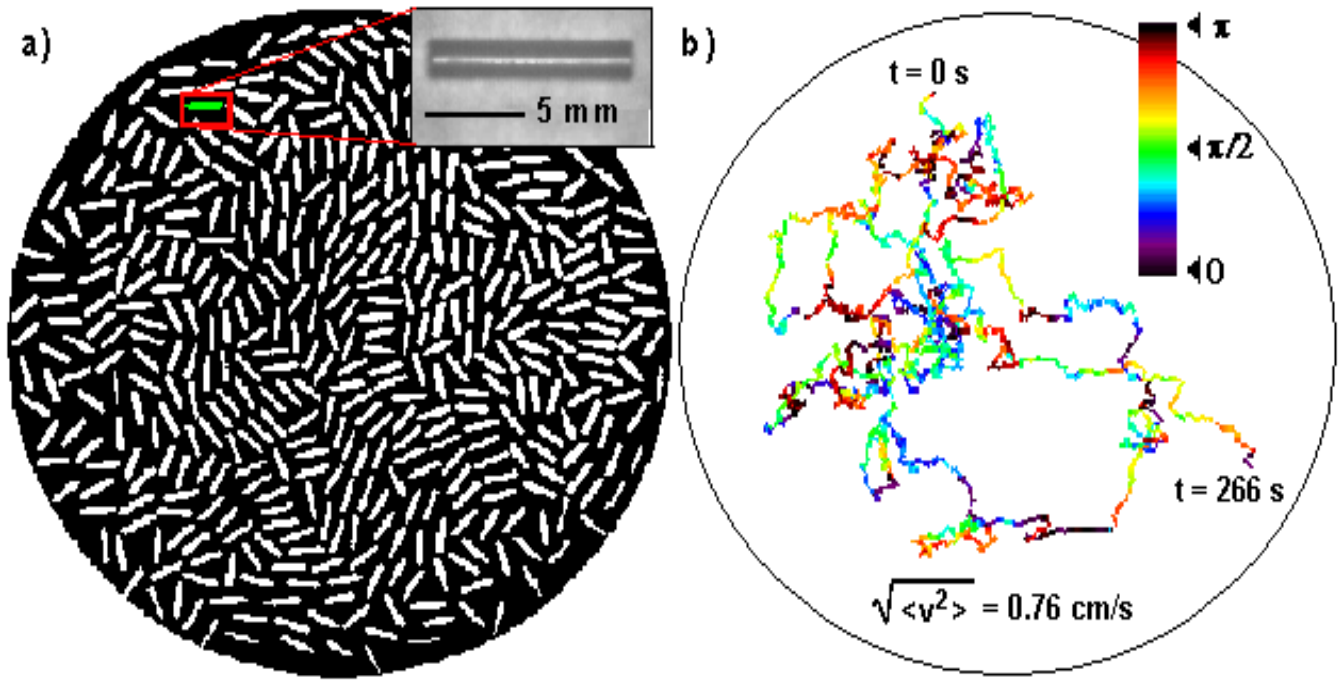}
\caption{(Color online) A monolayer of gas-fluidized bipolar rods.
The diameter of the system is $6.75''$. a) A frame capture of the
system from the video data after thresholding. The particles are
plastic dowel pins -- length $3/8''$, width $3/32''$, and mass
0.055 g -- occupying an area fraction of $42\%$. The call-out
shows an actual particle. b) A time-trace of one particle
(highlighted in the left image). The color code corresponds to
instantaneous orientation with respect to the horizontal axis of
the image.}\label{SYSTEM}
\end{figure*}

The system we study, depicted in Fig.~\ref{SYSTEM}(a), consists of
a monolayer of 435 cylindrical plastic dowel pins -- length $L =
0.95$ cm, diameter $d = 0.24$ cm, and mass $m = 0.055$ g --
fluidized by an upflow of air. The particular area fraction $\phi
=42\%$ ensures that the particles are uniformly distributed across
the system. The particle aspect ratio, $L/d = 4$, is chosen to
prevent any long-range ordering. Although an effective harmonic
potential exists due to interactions with the confining
walls~\cite{Durian2004,Ojha2005}, volume exclusion interactions
and the turbulent, chaotic mixing of individual particle wakes
overwhelms interactions with the boundaries and eliminates this
external potential. Excluded volume interactions also serve to
prevent unidirectional whirling of the particles.

In-plane motion of the rods is excited by a uniform upflow of air
at speed $U=370$~cm/s.  This is lower than the terminal free-fall
speed, so the rods do not levitate or fully lift off the plane. It
is also low enough that the rods do not
``chatter"~\cite{Abate2005}. Instead, one end of the rod lifts
slightly off the plane and causes the rod to scoot preferentially
in the direction of the tilt.  This is most evident when only a
few rods are present; here, at finite area fraction, the mean free
distance between rods is small.  In addition to scooting, the
upflow of air also induces random short-time motion.  Since the
Reynolds number based on rod length and air speed is large, $\sim
10^{3}$, there is turbulence in the form of irregular wakes shed
by the rods.  The shedding frequency $f_v$ is given by a universal
value of the Strouhal number: ${\rm St} = f_v L / U = 0.18$; a new
wake is generated every time the air flows a distance of about
$5L$~\cite{Antonia}.  Therefore, the rods experience a
corresponding random kicking force that fluctuates on a time scale
$ \tau_v = 1/f_v \approx 0.018$~s~\cite{Durian2004}.

The apparatus itself and fluidization method are identical to
those of Refs.~\cite{Durian2004,Ojha2005}. The apparatus is a
rectangular windbox, 1.5 $\times$ 1.5 $\times$ 4 ft.$^{3}$,
positioned upright. A circular testing sieve with mesh size 150
$\mu$m and diameter $30.5$ cm rests horizontally on top. To reduce
alignment of the particles with the wall, we place the particles
in a free-standing cylindrical insert, inner diameter of $17$ cm
and thickness of $0.32$ cm, at the center of the larger bed. A
blower attached to the windbox base provides vertical airflow
perpendicular to the sieve. The upper and lower halves of the
windbox are separated by a 1-inch-thick foam filter between two
perforated metal sheets to eliminate large-scale structures in the
airflow. The flow rate and its uniformity are confirmed via a
hot-wire anemometer. Raw video data of the fluidized particles is
captured for 10 minutes at 120 frames per second by a digital
camera mounted above the apparatus. Post-processing of the video
data is accomplished using LabVIEW.

Figure~\ref{SYSTEM}(b) shows a sample time-trace, $\{ x(t), y(t),
\theta (t) \}$, of a single particle. The color code denotes the
instantaneous orientation of the particle with respect to the
horizontal axis. The motion appears heterogeneous: we note several
long stretches where the particle is preferentially moving
parallel to its long axis. Apart from these, the wandering of the
particle is like that of an isotropic particle undergoing a random
walk. To obtain this time-trace, we convert each frame of the raw
data to binary as it is saved to video from buffer. Using
LabVIEW's ``IMAQ Particle Analysis" algorithm, we locate the
centroid of each particle in the thresholded image, imposing an
upper bound on the allowable area of a single particle, and
determine the orientation of each particle with respect to a fixed
horizontal axis. Any particles identified above the area bound
consist of two or more particles that have collided. In order to
distinguish the individual particles, a series of image processing
steps known as erosions, similar to Ref.~\cite{Abate2005}, is
carried out. Figure~\ref{SYSTEM}(a) shows the result of the
erosion process for one frame in which all particles have been
separated.

%For the collision of exactly two long particles that cannot be
%separated via erosion, we are still able to deduce their
%individual positions and orientations. We assume that the
%centers-of-mass of the individual rods are collinear with the
%center-of-mass of the collided pair, lying along the pair's axis
%of lowest moment of inertia, $\hat{x}_{\theta}$. We determine the
%center-of-mass coordinates and $\hat{x}_{\theta}$ for the pair and
%then calculate the moment of inertia, $I_{zz}$, about the pair
%center-of-mass. We assume that the moment of inertia about the
%center of an individual rod is that of a cylinder: $I_{rod} =
%\frac{1}{12}$$m(3r^{2}+l^{2})$, where $r$ is the radius, $l$ the
%length, and $m$ the area of an individual rod. We then use the
%parallel axis theorem to solve for the distance along
%$\hat{x}_{\theta}$ where the individual centers-of-mass are
%located: $d = \pm[(I_{rod}-I_{zz})/A]^{1/2}$, where A is the area
%of the pair.
%[ALSO NEED HOW YOU GET ORIENTATION]

Next, we link together particle positions and orientations from
frame to frame by finding the minimum displacement between two
center-of-mass positions in subsequent frames. To ensure correct
matches, we constrain this with a maximum allowable displacement
and rotation. The resulting time-traces are then smoothed by a
running average. We estimate error in the position and angle data
as $\sigma_{rms}$/$\sqrt{N}$, where $\sigma_{rms}$ is the rms
deviation of the raw data from the smoothed data and $N$ is the
number of frames in the smoothing window. This yields errors of 18
$\mu$m and 3.3 mrads in the position and angle data, respectively.
Finally, to further minimize any wall effects, we include only
those segments for which the particles are at least three particle
diameters away from the wall.

\section{Self-propulsion}\label{shorttime}
\begin{figure}
\includegraphics[width=3.00in]{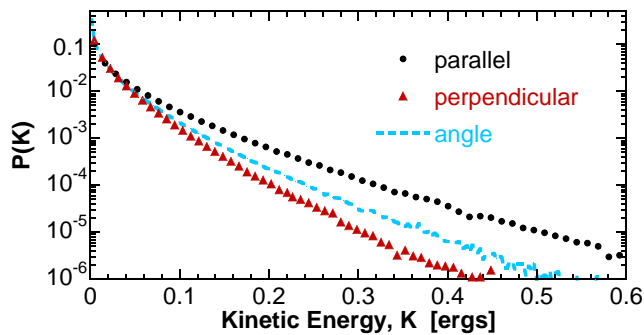}
\caption{(Color online) Distributions of kinetic energies,
K$_{\parallel}$ and K$_{\perp}$, for motion parallel to and
perpendicular to the particle's long axis and K$_{\theta}$, the
rotational kinetic energy. The average kinetic energy for each
components is $\langle K_{\parallel}\rangle$ = 0.021 ergs,
$\langle K_{\perp}\rangle$ = 0.012 ergs, and $\langle
K_{\theta}\rangle$ = 0.016 ergs. The kurtosis excess of the
velocity distributions are 0.5 for parallel, 0.4 for
perpendicular, and 0.6 for angle. A Gaussian distribution has a
kurtosis excess of 0.}\label{KEDIST}
\end{figure}

As noted in Section~\ref{experiment}, the sample time-trace,
Fig.~\ref{SYSTEM}(b), shows long stretches where particle
orientation is aligned with the direction of motion, indicating
scooting motion. To quantify this self-propelling behavior, we
obtain the distributions of kinetic energies for motion parallel
and perpendicular to the rod's long axis and for the rod's
orientation, shown in Fig.~\ref{KEDIST}. We immediately see that
equipartition of energy does not hold: at large energies, the
distribution of parallel kinetic energies is much greater than the
perpendicular distribution. The parallel component, with average
value $\langle K_{\parallel}\rangle$ =
1/2$m\langle$$v_{||}^{2}$$\rangle$ $\sim 0.021$ ergs, is roughly
twice as energetic as the perpendicular component, with $\langle
K_{\perp}\rangle$ = 1/2$m\langle$$v_{\perp}^{2}$$\rangle$ $\sim
0.012$ ergs. The rotational kinetic energy splits the difference
between the two. Its average value is roughly the average of these
two: $\langle K_{\theta}\rangle$ =
1/2$I\langle$$\omega^{2}$$\rangle$ $\sim 0.016$ ergs, where we
have used $I=(m/12)(3(d/2)^{2}+L^{2})$. This shows that the
gas-fluidized rods convert energy provided by the upflow of air
into in-plane motion, preferentially parallel to the long axis.
They do so at the expense of motion perpendicular to the rod's
long axis. Such microscopic dynamical anisotropy -- to which the
emergence of collective macroscopic behavior has been
attributed~\cite{Menon2006} -- is a universal feature of theories
and simulations of self-propelled particles, even when the
anisotropy is not specified as a part of the model \textit{a
priori}. We can rightly consider our gas-fluidized rods in the
context of self-propelled particles.

We also obtain the average self-propelling velocity,
$v_{\parallel} \sim 0.87$ cm/s, from $\langle
K_{\parallel}\rangle$. From this and the lab frame diffusion
coefficients $D_{x}$ and $D_{\theta}$ to be discussed in
Section~\ref{dynamics}, we calculate $\alpha =
v_{\parallel}^{2}/(D_{x} D_{\theta}) = 27$, a dimensionless
parameter that determines the extent to which self-propulsion
dominates over stochastic fluctuations~\cite{Marchetti2007}. For
values $\alpha > 4$, self-propulsion dominates. Thus, we are
clearly in a regime where self-propulsion effects will be readily
observable. Furthermore, we can quantify the spatial and temporal
extent of the `scooting' behavior by calculating an alignment
order parameter $m(t) = \cos[\theta(t) - \phi(t)]$, the cosine of
the angle between the particle's instantaneous orientation,
$\theta(t)$, and the direction of its instantaneous velocity,
$\phi(t)$. This quantity is 1 if a particle moves in the same
direction that it is pointing and 0 if it moves perpendicular to
its orientation. Here, $\langle m\rangle = 0.46$ and
$\sqrt{\langle m^{2} \rangle} =$ 0.75. From the autocorrelation
$\langle m(t)\cdot m(t+\tau)\rangle$, we extract a correlation
time of 6.4 s. We quantify the spatial extent of the stretches
from the value of the mean square parallel displacement at the
correlation time. This gives a displacement of 1.6 cm, roughly one
particle length per correlation time.

Self-propulsion is a strictly non-thermal phenomenon. The shape of
the energy distributions in Fig.~\ref{KEDIST} further indicates
the non-thermal character of our system. All of the distributions
deviate sharply from an exponential at small energies. This sharp
maximum has been observed for a single rod bouncing on a vibrated
surface~\cite{King2006}. This is due to the strong correlation
mentioned earlier between out-of-plane and in-plane motion.
Specifically, when the particle begins chattering out-of-plane,
its in-plane motion is significantly reduced, and vice versa. The
non-exponential form indicates that the velocity distributions are
non-Gaussian. We confirm this by calculating a non-Gaussian
parameter -- the kurtosis excess, $\langle v^{4}\rangle / \langle
v^{2}\rangle^{2} - 3$, which equals 0 for Gaussian distributions
-- of the velocity distributions. With kurtosis values of 0.6 for
the angle, 0.5 for the parallel, and 0.4 for the perpendicular,
none of the velocity distributions are Gaussian.

%\begin{figure}
%\includegraphics[width=3.00in]{vdist.eps}
%\caption{(Color online) Velocity distributions for the angle and
%body frame components. The solid curves are fits to a Gaussian.
%The inset shows the same distributions versus velocity-squared,
%emphasizing the long-time tails of the
%distributions.}\label{VDIST}
%\end{figure}

\section{Collisions}\label{collisions}
\begin{figure}
\includegraphics[width=3.00in]{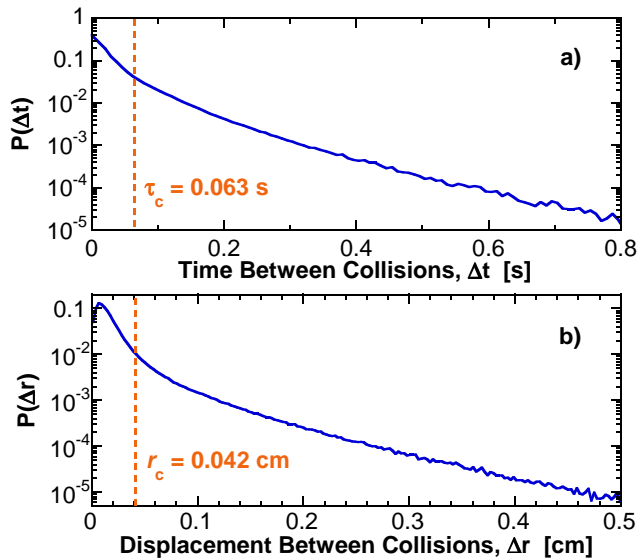}
\caption{(Color online) Probability distribution of a) times
between collisions and b) displacements between collisions. The
mean free time, $\tau_{c} \sim 0.063$ s, and mean free path,
$r_{c} \sim 0.042$ cm, are labeled.\label{MFT}}
\end{figure}

The anisotropy of particle shape gives rise to self-propelled
behavior and also alters the excluded volume interaction between
particles. As a first step toward understanding these steric
interactions, we compile binary collision statistics. We generate
distributions of free times and free paths between collisions,
shown in Figure~\ref{MFT}, using a graphical approach. We create a
particle template that is the size of a grain had it not undergone
any erosions or thresholding during data postprocessing. Then, we
reconstruct each frame of the video by overlaying the template at
the center-of-mass of each actual particle and rotating it by the
particle's orientation. By detecting particle overlap, we
determine whether a given particle has collided with any neighbors
in a given frame. We then extract the time increment, $\Delta$$t$,
and displacement, $\Delta r$, between collisions for a given
particle and compile them as probability distributions of free
times, Fig.~\ref{MFT}(a), and free paths, Fig.~\ref{MFT}(b).

Although the tails of the distributions in Fig.~\ref{MFT} behave
exponentially, there is substantial deviation at short times and
paths. As such, we obtain well-defined mean values by taking the
average of all the data rather than from an exponential fit. This
yields a mean free time between collisions, $\tau_{c} \sim 0.063$
s, and a mean free path, $r_{c} \sim 0.042$ cm, marked as the
dashed vertical lines in the figure. The ratio, $r_{c}/\tau_{c} =
0.67$ cm/s, is near the rms velocity, 0.76 cm/s. On subsequent
dynamics plots, we mark the collision time with an orange dashed
vertical line.

For thermal, isotropic systems, collisions are independent events;
we expect exponential distributions for free paths and free times.
The deviation of the distributions in Fig.~\ref{MFT} from
exponential behavior at short times and paths shape further
indicates non-thermal behavior. Distributions similar to ours were
observed for spherical grains vibrated on an inclined
plane~\cite{Blair2003}. There, the sharp maximum was attributed to
inelastic clustering. Here, it may be attributable to local
alignment -- which may make a subsequent collision dependent on
prior collisions --  or to dynamical anisotropy. That is, a
particle may be more likely to collide at short times as it
propels through its environment.

\section{Lab Frame Dynamics}\label{dynamics}
\begin{figure}
\includegraphics[width=3.00in]{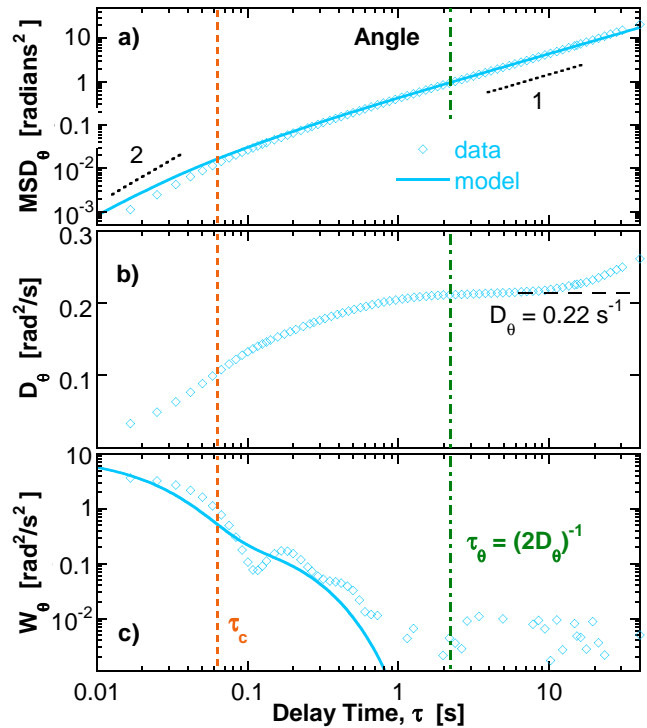}
\caption{(Color online) Dynamics of rod orientation: a) mean
square angular displacement MSD$_{\theta}$, b) angular diffusion
coefficient $D_{\theta}=$ MSD$_{\theta}/(2\tau)$, and c) angular
velocity autocorrelation function. The line with slope 1 in a)
corresponds to late-time diffusive motion; the line with slope 2
shows short-time ballistic motion. The horizontal line in b) is
the rotational diffusion coefficient, $D_{\theta}$= 0.22 s$^{-1}$.
The dashed vertical line (orange) marks the collision time,
$\tau_{c}$; the dash-dot line (green) marks the directional memory
time, $\tau_{\theta} = (2D_{\theta})^{-1}$. The solid curves are a
Langevin model using a non-thermal noise source, given by
Eqs.~(\ref{rotvaf}) and (\ref{rotmsd}).\label{ROTFRAME}}
\end{figure}

Despite being driven far from equilibrium, one or two
gas-fluidized spheres act as Brownian particles in a harmonic
trap~\cite{Durian2004}. However, in Section~\ref{shorttime}, we
showed that a rod moves more energetically parallel to its long
axis and this thermal analogy subsequently fails. Here, we further
ask how does self-propulsion alter the dynamics of a gas-fluidized
rod? We begin by calculating single-particle dynamics in the lab
frame where particle motion is characterized by both angular and
translational displacements. The lab frame axes, $x$ and $y$,
correspond to the horizontal and vertical axes of the raw images;
angular orientation is measured counterclockwise with respect to
$+x$. Recall from Section~\ref{experiment} that we only analyze
data for which a particle is within 3 particle diameters from the
wall, effectively breaking a single time-trace into many shorter
time-traces. A typical particle moves across the system in less
than a minute. Therefore, dynamical quantities, such as
MSD$_{\theta}$, are truncated for delay times greater than 40 s.

Figure~\ref{ROTFRAME}(a) shows the rotational mean square
displacement, MSD$_{\theta}(\tau) = \langle[\theta(t+ \tau)-
\theta(t)]^{2}\rangle$, as a function of delay time $\tau$.
%Our shortest time resolution is smaller than the
%momentum relaxation time for our particles, $D_{\theta}/\langle
%\omega^{2}\rangle \sim 0.054$ s, and
We observe ballistic behavior ($\propto$ $\tau^{2}$) at the
shortest time scales and a crossover to diffusive behavior
($\propto$ $\tau$) at long times.

Rotational diffusion, characterized by the rotational diffusion
coefficient $D_{\theta}(\tau) =$ MSD$_{\theta}/(2\tau)$ shown in
Fig.~\ref{ROTFRAME}(b), sets a `directional memory' timescale
$\tau_{\theta}= (2D_{\theta})^{-1}$. At times greater than
$\tau_{\theta}$, a particle will have forgotten its initial
direction and all directions become equal. The long-time value of
the angular diffusion coefficient, $D_{\theta}=0.22$ s$^{-1}$, is
obtained from the plateau in Fig.~\ref{ROTFRAME}(b), giving the
value $\tau_{\theta} \sim 2.27$ s, shown as the vertical (green)
dashed line in Fig.~\ref{ROTFRAME}.

Chiral particles subject to external forces, such as those arising
from the vibration of a substrate~\cite{GollubChiral} or from air
flowing past them as in the current experiment, spin in a
preferred direction determined by the sign of their chirality.
Small manufacturing defects impart chirality with a random sign
and magnitude to some of the rods in this experiment. Some of the
long-time rise in $D_{\theta}(\tau)$ is due to the spinning of
some particles throughout the entire data set. If these particles
are removed, an increase in slope is still detected, indicating
heterogeneous dynamics for non-whirling particles. Just as the
time-trace in Fig.~\ref{SYSTEM}(b) shows regions where particle
scooting is intermittent with thermal-like wiggling, plots of
$\theta (t)$ show regions of fluctuating motion intermittent with
rapid rotation. We stress that, if we exclude whirlers, the
long-time value of $D_{\theta}$, and thus the timescale
$\tau_{\theta}$, does not change significantly.

The angular velocity autocorrelation, $W_{\theta}(\tau) =\langle
\omega(t+\tau)\cdot \omega(t)\rangle$, plotted in
Fig.~\ref{ROTFRAME}(c), shows a two-step decay, characterized by a
large, positive rebound near $\tau_{c}$ due to collisions,
followed by small oscillations at longer times before noise
dominates. The behavior of $W_{\theta}(\tau)$ at $\tau_{c}$ allows
us to deduce the effect that collisions have on the particle.
Typically, a collision results in a negative rebound: the particle
recoils in a direction opposite to its incident direction. The
large, positive rebound in $W_{\theta}(\tau$) suggests that the
nature of our collisions is to re-align the particle as to its
initial direction.

\begin{figure}
\includegraphics[width=3.00in]{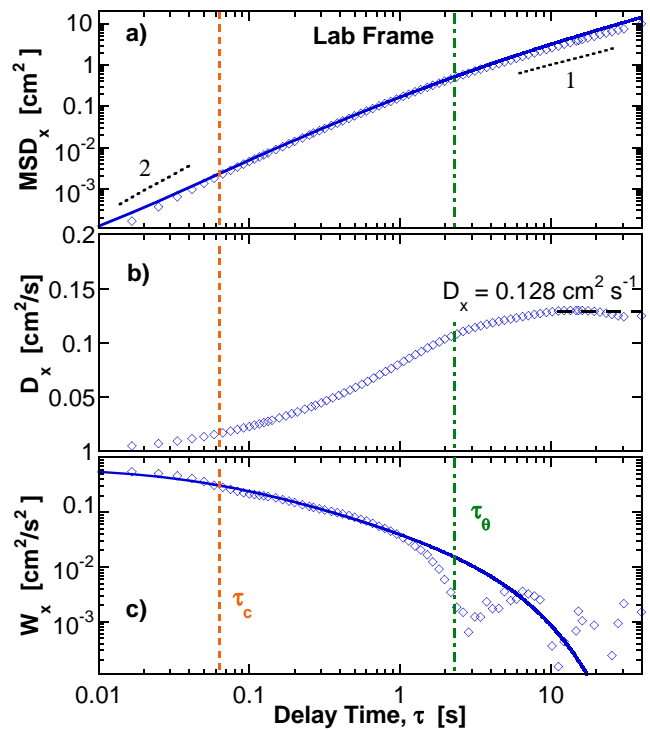}
\caption{(Color online) Dynamics of rod position in the lab frame:
a) mean square displacement MSD$_{x}$, b) diffusion coefficient
$D_{x}=$ MSD$_{x}/(2\tau)$, and c) velocity autocorrelation
function. The horizontal line in b) shows the long-time value of
the diffusion coefficient, $D_{x} = 0.128$ cm$\cdot$s$^{-1}$. The
dashed vertical line (orange) marks the collision time,
$\tau_{c}$; the dash-dot line (green) marks the directional memory
time, $\tau_{\theta} = (2D_{\theta})^{-1}$. The solid curves are a
Langevin model using a non-thermal noise source, given by the
first term in Eq.~\ref{falfvaf}.\label{LABFRAME}}
\end{figure}

We compute the translational dynamics in the lab frame, shown in
Fig.~\ref{LABFRAME}. There is no discernable difference between
data along $x$ and $y$; only data along $x$ is plotted. Figure
~\ref{LABFRAME}(a) shows the lab frame mean square displacement,
MSD$_{x}(\tau) = \langle[x(t+ \tau)- x(t)]^{2}\rangle$. The
behavior is ballistic at short times and becomes diffusive at
longer times. This is confirmed in Fig.~\ref{LABFRAME}(b): the lab
frame diffusion coefficient, $D_{x}(\tau) =$ MSD$_{x}/(2\tau)$,
approaches its long-time value, $D_{x}=0.128$ cm/s, at about 10 s.
The lab frame velocity autocorrelation function, $W_{x}(\tau)$,
shown in Fig.~\ref{LABFRAME}(c), has slower-than-exponential decay
with a small wiggle at $\tau_{c}$. Long-time statistics are poorer
and the rebound at $\tau_{\theta}$ may or may not be real.

Equipartition of energy holds roughly between the lab frame and
the angle. From $W_{x}(0)$ and $W_{\theta}(0)$, we find that
$(1/2) m\langle v_{x}^{2} \rangle \sim$ 0.017 ergs, and
$(1/2)I\langle$$\omega^{2}$$\rangle$ $\sim$ 0.016 ergs.
Remarkably, despite the self-propelled nature of the particles,
averaging over all possible orientations results in bulk behavior
that is more or less thermal in appearance. This is consistent
with the results in Fig.~\ref{KEDIST}; the distribution of kinetic
energies in the lab frame is nearly identical to that of the angle
and the average of the parallel and perpendicular distributions.

\section{Coupling Rotation and Translation}\label{coupling}

In this section, we ask how do rotation and translation couple for
a self-propelling particle? To explicitly visualize
rotational-translational coupling, we calculate the dynamics in a
`fixed-angle' lab frame with axes $\bar{x}$ and $\bar{y}$. We
construct this frame by rotating the coordinates of an entire time
trace by the initial particle orientation so that $\bar{x}$ and
$\bar{y}$ are, respectively, parallel and perpendicular to the
initial direction of the long axis of the particle. The axes then
remain fixed in time. This is equivalent to setting the initial
orientation of all particles to $\theta_{0}=0$.

For comparison purposes, it is instructive to review how rotation
and translation couple for a Brownian
particle~\cite{Perrin1934,Perrin1936,Yodh2006}. If a particle is
not allowed to rotate, translational motion is characterized by
anisotropic diffusion -- with two diffusion coefficients,
$D_{\parallel}$ and $D_{\perp}$ -- for displacements parallel and
perpendicular to the particle's long axis. The two components, it
should be noted, become diffusive on the same timescale. If the
particle is allowed to rotate, this anisotropic diffusion regime
will cross over to isotropic diffusion characterized by a single
diffusion coefficient, $D_{x}= (1/2)(D_{\parallel}+ D_{\perp}$).
The crossover timescale is the same `directional memory' timescale
discussed earlier in Section~\ref{dynamics}, $\tau_{\theta}=
(2D_{\theta})^{-1}$. Thus, at times longer than $\tau_{\theta}$,
the fixed-angle lab frame axes will become random and the dynamics
along $\bar{x}$ and $\bar{y}$ will become equivalent to the
conventional lab frame, $x$ and $y$.

\begin{figure}
\includegraphics[width=3.00in]{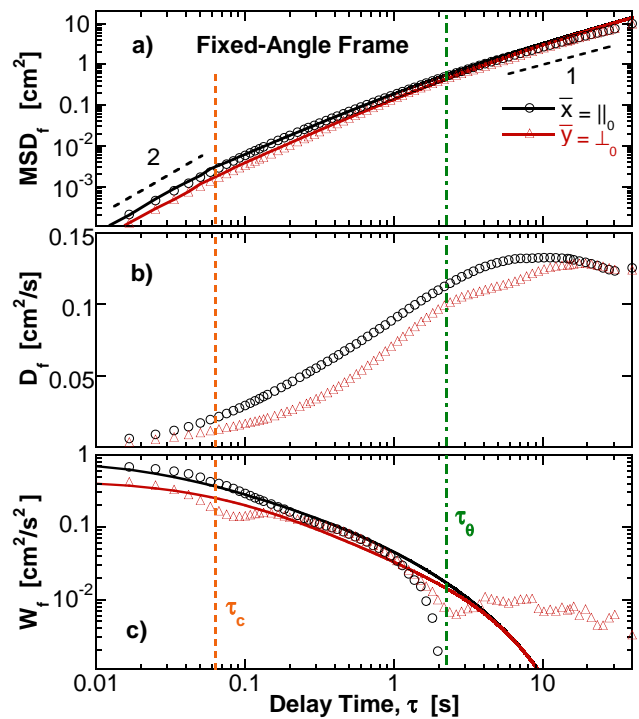}
\caption{(Color online) Dynamics of rod position in the
fixed-angle lab frame: a) mean square displacement, b) diffusion
coefficient $D_{f}=$ MSD$_{f}/(2\tau)$, and c) velocity
autocorrelation function. At t=0, $\tilde{x}$ is aligned parallel
to the particle's long axis. The dashed vertical line (orange)
marks the collision time, $\tau_{c}$; the dash-dot line (green)
marks the directional memory time, $\tau_{\theta} =
(2D_{\theta})^{-1}$. The solid curves are a Langevin model using
non-thermal noise sources, given by
Eq.~(\ref{falfvaf}).\label{FALF}}
\end{figure}

The fixed-angle lab frame mean square displacements
MSD$_{\bar{x}}(\tau)$ and MSD$_{\bar{y}}(\tau)$,
Fig.~\ref{FALF}(a), are both ballistic at short times. This
short-time behavior confirms the discussion in
Section~\ref{shorttime} that equipartition of energy does not
hold, with MSD$_{\bar{x}}(\tau)$ twice as large as
MSD$_{\bar{y}}(\tau)$. After $\tau_{\theta}$, the fixed-angle MSDs
become diffusive and eventually converge, showing that the
coordinate axes are randomized and the particle has forgotten its
initial direction. This indicates coupling between rotation and
translation in our system. However, examining Fig.~\ref{FALF}(b),
we do not observe a fully-developed anisotropic diffusion regime.
Instead, following the initial anisotropic ballistic regime,
$\bar{D}_{x}$ indicates diffusion at approximately 6 s but
$\bar{D}_{y}$ does not become diffusive until about 20 s.
Isotropic diffusion occurs near 20 s, when the two diffusion
coefficients converge.

We see that coupling for self-propelled particles is different
than Brownian in two significant ways. First, the
intermediate-time dynamics show that the two components become
diffusive on different timescales. Recalling the
ballistic-to-diffusive timescale for the conventional
orientation-averaged lab frame, $\bar{x}$ becomes diffusive sooner
and $\bar{y}$ later. This is most likely because the parallel
component is more energetic than the perpendicular component;
thus, it takes longer for the $\bar{y}$ component to `catch up'
and become equal to $\bar{x}$. Secondly, isotropic diffusion
occurs at a timescale an order of magnitude larger than
$\tau_{\theta}$. Thus, we think of self-propulsion as a `memory
enhancement' effect. Isotropic diffusion occurs much later because
self-propulsion, in effect, allows the particle to remember its
initial direction for a longer time.

\section{Body Frame Dynamics}\label{body}
\begin{figure}
\includegraphics[width=3.00in]{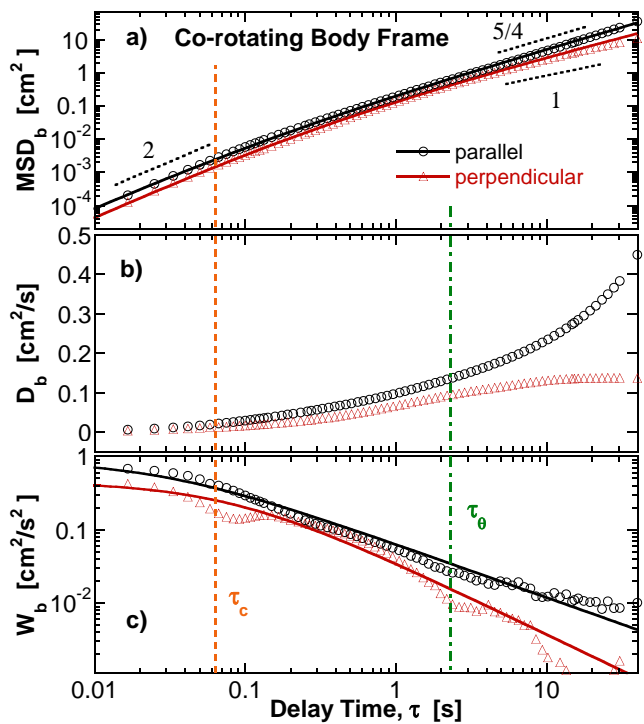}
\caption{(Color online) Dynamics of rod position in the body
frame: a) mean square displacement MSD$_{b}$, b) diffusion
coefficient $D_{b}=$ MSD$_{b}/(2\tau)$, and c) velocity
autocorrelation function. The co-rotating body frame axes are
always either parallel or perpendicular to the particle's long
axis. The line with slope $5/4$ in a) corresponds to
superdiffusive motion. The dashed vertical line (orange) marks the
collision time, $\tau_{c}$; the dash-dot line (green) marks the
directional memory time, $\tau_{\theta} = (2D_{\theta})^{-1}$. The
solid curves are a Langevin model using a non-thermal noise
source, given by Eqs.~(\ref{bfpl}) and
(\ref{bfmsd}).\label{BODYFRAME}}
\end{figure}

While the fixed-angle lab frame is useful for illustrating the
effects of self-propulsion at short and intermediate times, it is
useful to consider yet another reference frame to capture the
effects at longer times. The body frame is a set of axes,
$\tilde{x}$ and $\tilde{y}$, which are re-oriented at each time
step to coincide with the long and short dimensions of each rod,
respectively. The individual displacements are then summed up
successively to form a set of time-traces,
$\{\tilde{x}(t),\tilde{y}(t)\}$. Here, $\theta (t) = 0$ at all
times and there is no coupling in this frame.

At short times, the body frame MSDs in Fig.~\ref{BODYFRAME}(a) are
identical to the fixed-angle lab frame MSDs of Fig.~\ref{FALF}(a).
At long times, MSD$_{\tilde{y}}$ crosses over to diffusive
behavior whereas MSD$_{\tilde{x}}$ becomes superdiffusive
($\propto\tau^{5/4}$). Self-propulsion gives rise to enhanced
diffusion for translations along the particle's long axis. As seen
in Fig.~\ref{BODYFRAME}(b), the perpendicular diffusion
coefficient, $D_{\perp}$, has reached its long-time value while
$D_{\parallel}$ continues to increase.

The body-frame velocity autocorrelation functions,
$W_{\parallel}(\tau$) and $W_{\perp}(\tau$), shown in
Fig.~\ref{BODYFRAME}(c), reveal the long memory of the particle
motion. The perpendicular component has a slow decay that exhibits
the same rebounding features at $\tau_{c}$ and $\tau_{\theta}$ as
seen in $W_{\theta}(\tau$) and $W_{x}(\tau)$. The parallel
component shows a smooth, slower algebraic decay for the entire
run indicating remarkably long-lived velocity correlations. This
is consistent with the long, unbroken stretches of scooting motion
seen in the time-trace image, Fig.~\ref{SYSTEM}(b).

\section{Model}\label{langevin}

Collections of self-propelled particles are typically modeled in
one of two ways: establishing a minimal set of
rules~\cite{Vicsek1995,Chate2004} or writing hydrodynamic
equations, including all terms consistent with relevant
symmetries~\cite{Tu1995,Tu1998,Ramaswamy2005}. Self-propulsion is
usually included as a phenomenological parameter. In keeping with
such minimal models, we describe our system with a Langevin
formalism constructed along the lines of that for an anisotropic
Brownian particle. We make a single modification: rather than
write a self-propelling force, we implicitly include
self-propulsion with non-thermal noise terms.
%Not only will this formalism demonstrate to what extent the
%rotational-translational coupling is like that of a Brownian
%particle, it will also demonstrate whether self-propulsion must be
%more explicitly taken into account.
We begin by constructing Langevin equations for the three
independent degrees of freedom -- the angle and the two body frame
components.

The effective harmonic potential of the bed has been eliminated
and there are no externally imposed forces or torques. The
remaining forces -- arising from interparticle collisions,
interactions with the substrate, and hydrodynamic interactions
with wakes -- can be considered as noise. Thus, our Langevin
equations simply relate time derivatives of angle and displacement
to random-noise torques and forces. The equation for the
orientation of the rod, $\theta(t)$, is
\begin{equation}\label{rotlang}
    \frac{d\theta}{dt} \equiv \omega (t)= \zeta_{\theta}(t)
\end{equation}
where $\zeta_{\theta}(t)$ is Gaussian angular noise with zero
average and variance

\begin{equation}\label{rotlangnoise}
    \langle\zeta_{\theta}(t)\zeta_{\theta}(t')\rangle = \langle\omega(t)\omega(t')\rangle \equiv
    W_{\theta}(\tau).
\end{equation}
From Eq.~(\ref{rotlang}), $W_{\theta}(\tau)$ is the velocity
autocorrelation function of Fig.~\ref{ROTFRAME}(c).

The equations for displacement can be expressed in the lab frame
\textbf{x$_{lab}$}$(t) = \{ x(t), y(t)\}$ or in the body frame,
\textbf{$\tilde{x}$}$(t) = \{ \tilde{x}(t), \tilde{x}(t)\}$.
%, using the rotation matrix
%\begin{equation}\label{rotmat}
%    R_{ij}[\theta(t)]=\left(%
%\begin{array}{cc}
%  \cos \theta(t) & \sin \theta(t) \\
%  -\sin \theta(t) & \cos \theta(t) \\
%\end{array}%
%\right).
%\end{equation}
In the body frame, $\tilde{x}$(t) and $\tilde{y}$(t) decouple:
\begin{equation}\label{bflang}
    \frac{d\tilde{x}}{dt}\equiv \tilde{v}_{x}(t) = \zeta_{\parallel}(t),
\end{equation}
\begin{equation}\label{bflang2}
    \frac{d\tilde{y}}{dt}\equiv \tilde{v}_{y}(t) = \zeta_{\perp}(t),
\end{equation}
where $\zeta_{\parallel}(t)$ and $\zeta_{\perp}(t)$ are Gaussian
random noises with zero mean and variance
\begin{equation}\label{bflangnoise}
    \langle \zeta_{\parallel}(t) \zeta_{\parallel}(t')\rangle = \langle v_{\parallel}(t)v_{\parallel}(t')\rangle \equiv
    W_{\parallel}(\tau)
\end{equation}
\begin{equation}\label{bflangnoise2}
    \langle \zeta_{\perp}(t) \zeta_{\perp}(t')\rangle = \langle v_{\perp}(t)v_{\perp}(t')\rangle \equiv
    W_{\perp}(\tau).
\end{equation}
The noises $W_{\perp}(\tau)$ and $W_{\parallel}(\tau)$ are the
body frame velocity autocorrelation functions of
Fig.~\ref{BODYFRAME}(c).
%We justify the latter by recalling that, because
%fluidized rods move by scooting, the particle only interacts with
%the surface by colliding with it.

The Langevin equations derived above apply to both equilibrium and
non-equilibrium systems. In equilibrium systems, the noise
fluctuations of Eqs.~(\ref{rotlangnoise}),~(\ref{bflangnoise}),
and~(\ref{bflangnoise2}) are determined by the
fluctuation-dissipation theorem. They adopt white-noise forms when
rotational and translational friction coefficients do not exhibit
memory effects. Non-equilibrium systems are not restricted by the
fluctuation-dissipation theorem. Here, we use the experimental
forms of the angular and translational velocity autocorrelation
functions to set $W_{\theta}(\tau)$, $W_{\perp}(\tau)$, and
$W_{\parallel}(\tau)$. Since these quantities contain all the
information about self-propulsion, we are able to implicitly
include self-propulsion in our model via non-thermal noise. We
will then be able to test whether modeling self-propulsion as
non-thermal noise is sufficient in lieu of specifying an actual
force acting along the long axis of the particle.

The form of the angular noise $\zeta_{\theta}(t)$ can be
determined from the two-step decay of $W_{\theta}(\tau$),
Fig.~\ref{ROTFRAME}(c). This suggests the sum of two exponentials:
\begin{equation}\label{rotvaf}
    W_{\theta}(\tau)=D_{\theta}(\frac{a}{\tau_{1}}e^{-|\tau|/\tau_{1}}
    + \frac{1-a}{\tau_{2}}e^{-|\tau|/\tau_{2}})
\end{equation}
where $a$ is a real number between 0 and 1. The fit to
Eq.~(\ref{rotvaf}), shown as the solid curve in
Fig.~\ref{ROTFRAME}(c), was made by constraining $D_{\theta}$ to
equal its long-time value, 0.22 s$^{-1}$, rather than using it as
a fitting parameter. Although unable to capture the sharp rebound
caused by collisions, Eq.~(\ref{rotvaf}) provides a good fit to
$W_{\theta}(\tau)$. We extract two correlation times: $\tau_{1}$=
0.018 $\pm$ 0.005 s and $\tau_{2}$ = 0.11 $\pm$ 0.01 s. The
smaller correlation time is identical to the vortex shedding
timescale $\tau_\nu$ calculated in Section~\ref{experiment}. The
value of $\tau_{2}$ is roughly the collision time.

We also calculate the analytical form of the mean square angular
displacement:
\begin{equation}\label{rotmsd}
    \langle(\Delta\theta)^{2}\rangle = 2D_{\theta}
    [aS(t-t',\tau_{1})+(1-a)S(t-t',\tau_{2})]
\end{equation}
where we define
\begin{equation}\label{S}
    S(t,\tau)=|t|-\tau(1-e^{-|t|/\tau}).
\end{equation}
Using the fit values from Eq.~(\ref{rotvaf}), we plot
Eq.~(\ref{rotmsd}) as the solid curve in Fig.~\ref{ROTFRAME}(a).
As a consequence of constraining the value of $D_{\theta}$ when
fitting to Eq.~(\ref{rotvaf}), the result overestimates the value
of the MSD at short times.

The form of the body frame noise, $\zeta_{\parallel}(t)$ and
$\zeta_{\perp}(t)$, is obtained from $W_{\parallel}(\tau$) and
$W_{\perp}(\tau$), Fig.~\ref{BODYFRAME}(c), respectively. The slow
decay suggests a power law form:
\begin{equation}\label{bfpl}
    W_{i}(\tau)=\frac{A_{i}}{(1+b_{i}\tau)^{\alpha_{i}}}.
\end{equation}
The solid curves in Fig.~\ref{BODYFRAME}(c) are fits to
Eq.~(\ref{bfpl}), yielding exponents of $\alpha$$_{\perp}$ = 0.99
$\pm$ 0.04 and $\alpha$$_{\parallel}$ = 0.73 $\pm$ 0.02. These
terms implicitly include the two non-thermal effects of
self-propulsion. First, the magnitudes, $A_{i}$, contain
information about the energy gap between the two components.
Secondly, the power law exponents incorporate the extended memory
effect of self-propulsion. We note that a power law decay,
$\tau^{-d/2}$, is expected for particles suspended in a fluid due
to diffusive transport of momentum through the surrounding
fluid~\cite{Alder1967}. Although this is not the case for our
study, we highlight it as a potential analogy: particles in a
viscous medium and self-propelling particles both exhibit velocity
autocorrelations with extended memory effects. This form is also
able to capture the superdiffusive behavior of
MSD$_{\parallel}(\tau)$. We obtain the following expression for
the body frame MSDs:

\begin{eqnarray*}
        \langle(\Delta \tilde{x}_{i})^{2}\rangle=\frac{2A_{i}}{(1-\alpha_{i})(2-\alpha_{i})b_{i}^{2}}[(a_{i}+b_{i}\tau)^{2-\alpha_{i}}-a_{i}^{2-\alpha_{i}}]
\end{eqnarray*}
\begin{eqnarray}\label{bfmsd}
    -\frac{2A_{i}}{(1-\alpha_{i})b_{i}}a_{i}^{1-\alpha_{i}}\tau.
\end{eqnarray}
This form shows a crossover from $\tau^{2}$ at short times to
$\tau^{2-\alpha_{i}}$ at long times. Using the parameters obtained
from the fit to Eq.~(\ref{bfpl}), we plot the functional forms
given by Eq.~(\ref{bfmsd}) as the solid lines in
Fig.~\ref{BODYFRAME}(a).

We now have enough information to construct the coupled
fixed-angle lab frame dynamics. Writing the velocity in the
fixed-angle lab frame in terms of the body frame velocity,
\begin{equation}\label{vFALF}
    \bar{v}_{k}(t)=R_{kl}^{-1}(\theta (t))\tilde{v}_{l}(t)
\end{equation}
we obtain for the fixed-angle lab frame velocity autocorrelation
function:
\begin{eqnarray}\label{falfvaf}
    W_{f}(t,t')=\frac{1}{2}[W_{\parallel}(\tau)+W_{\perp}(\tau)]e^{-\langle(\Delta\theta)^{2}\rangle/2}\delta_{ij}\nonumber\\
    +\frac{1}{2}[W_{\parallel}(\tau)-W_{\perp}(\tau)]M_{ij}(\theta(0))e^{-\langle(\theta(t')+\theta(t))^{2}\rangle/2},
\end{eqnarray}
where $\theta(0) = 0$ and
\begin{equation}\label{Mij}
    M_{ij} =\left(%
\begin{array}{cc}
  \cos 2\theta(t) & \sin 2\theta(t) \\
  \sin 2\theta(t) & -\cos 2\theta(t) \\
\end{array}%
\right).
\end{equation}

We use the fitting parameters obtained from our fits to
Eqs.~(\ref{rotvaf}) and (\ref{bfpl}) to generate
Eq.~(\ref{falfvaf}), plotted as the solid curves in
Fig.~\ref{FALF}(c). The model shows that $W_{\tilde{x}}(\tau)$ and
$W_{\tilde{y}}(\tau)$ converge at $\tau_{\theta}$, consistent with
a crossover in the data.

We then numerically integrate $W_{f}$(t,t') according to
\begin{equation}\label{vaftomsd}
 \langle[\Delta \bar{x}(t)]^{2}\rangle = \int^{t}_{0}dt_{1}\int^{t}_{0}dt_{2} W_{f}(t_{1},t_{2})
\end{equation}
to obtain the solid curves in Fig.~\ref{FALF}(a). The components
of the model converge on the same timescale as the data. We see
that modelling self-propulsion as an external noise source with
long-lived correlations is sufficient to reproduce
rotational-translational coupling. Our model suggests that a
phenomenological way to think about collections of self-propelling
particles is in terms of enhanced memory effects rather than
explicitly detailing novel forces and torques in microscopic
equations.

The model also reproduces the angle-averaged lab frame dynamics
well (Fig.~\ref{LABFRAME}). Averaging our expressions over all
initial angles eliminates the second term in Eq.~(\ref{falfvaf});
the result is plotted in Fig.~\ref{LABFRAME}(c) as the solid
curve. We integrate according to Eq.~(\ref{vaftomsd}) to obtain
the solid curve in Fig.~\ref{LABFRAME}(a). The model describes the
angle-averaged velocity correlations very well out to 1 s, after
which the data falls off more rapidly. The good agreement with the
fit here confirms the bulk thermal behavior of the collection of
self-propelling rods.

\section{Conclusion}
We have investigated the dynamics of gas-fluidized rods. Particle
shape anisotropy leads to dynamical anisotropy, characterized by
preferential motion parallel to the particle's long axis. Ours is
a model system -- with the advantage of a temporally and spatially
homogeneous driving method -- to further investigate universal
phenomena predicted for collections of self-propelled particles.

In this report, we compared the coupling of rotation and
translation couple for a self-propelled particle to that of an
anisotropic Brownian particle. A modified Langevin formalism
implicitly specifying self-propulsion via non-thermal noise
describes the dynamics data well, capturing
rotational-translational coupling at the correct timescale.
Despite the energy gap between the parallel and perpendicular
components, the model was able to reproduce the loss of
directional memory at long times. Furthermore, despite the
non-thermal behavior of individual particles, the bulk
angle-averaged behavior is nearly thermal.

Future work will continue to explore phase space in the interest
of observing collective behavior and spontaneous symmetry breaking
for denser collections of both bipolar and polar self-propelled
particles. We are interested in whether we can induce collective
macroscopic behavior by manipulating the boundaries of the system
as well. %We
%are interested in the extent to which preferred motion independent
%of particle geometry gives rise to coherent behavior.
We are currently working on characterizing compression waves that
propagate through denser collections of gas-fluidized rods. We
hope that, through comparison with theoretical models and recent
vibrated-bed experiments, our system will further shed light on
universal behavior of collections of self-propelled particles.

\begin{acknowledgments}
We thank Adam Abate for helpful discussion and for programming
assistance. This work was supported by the NSF through grant
DMR-0704147.
\end{acknowledgments}

\bibliography{RodsRefs}

\begin{thebibliography}{30}
\expandafter\ifx\csname natexlab\endcsname\relax\def\natexlab#1{#1}\fi
\expandafter\ifx\csname bibnamefont\endcsname\relax
  \def\bibnamefont#1{#1}\fi
\expandafter\ifx\csname bibfnamefont\endcsname\relax
  \def\bibfnamefont#1{#1}\fi
\expandafter\ifx\csname citenamefont\endcsname\relax
  \def\citenamefont#1{#1}\fi
\expandafter\ifx\csname url\endcsname\relax
  \def\url#1{\texttt{#1}}\fi
\expandafter\ifx\csname urlprefix\endcsname\relax\def\urlprefix{URL }\fi
\providecommand{\bibinfo}[2]{#2}
\providecommand{\eprint}[2][]{\url{#2}}

\bibitem[{\citenamefont{Vicsek et~al.}(1995)\citenamefont{Vicsek, Czirok,
  Ben-Jacob, Cohen, and Shochet}}]{Vicsek1995}
\bibinfo{author}{\bibfnamefont{T.}~\bibnamefont{Vicsek}},
  \bibinfo{author}{\bibfnamefont{A.}~\bibnamefont{Czirok}},
  \bibinfo{author}{\bibfnamefont{E.}~\bibnamefont{Ben-Jacob}},
  \bibinfo{author}{\bibfnamefont{I.}~\bibnamefont{Cohen}}, \bibnamefont{and}
  \bibinfo{author}{\bibfnamefont{O.}~\bibnamefont{Shochet}},
  \bibinfo{journal}{Phys. Rev. Lett.} \textbf{\bibinfo{volume}{75}},
  \bibinfo{pages}{1226} (\bibinfo{year}{1995}).

\bibitem[{\citenamefont{Gregoire and Chate}(2004)}]{Chate2004}
\bibinfo{author}{\bibfnamefont{G.}~\bibnamefont{Gregoire}} \bibnamefont{and}
  \bibinfo{author}{\bibfnamefont{H.}~\bibnamefont{Chate}},
  \bibinfo{journal}{Phys. Rev. Lett.} \textbf{\bibinfo{volume}{92}},
  \bibinfo{pages}{025702} (\bibinfo{year}{2004}).

\bibitem[{\citenamefont{Toner and Tu}(1995)}]{Tu1995}
\bibinfo{author}{\bibfnamefont{J.}~\bibnamefont{Toner}} \bibnamefont{and}
  \bibinfo{author}{\bibfnamefont{Y.}~\bibnamefont{Tu}}, \bibinfo{journal}{Phys.
  Rev. Lett.} \textbf{\bibinfo{volume}{75}}, \bibinfo{pages}{4326}
  (\bibinfo{year}{1995}).

\bibitem[{\citenamefont{Toner and Tu}(1998)}]{Tu1998}
\bibinfo{author}{\bibfnamefont{J.}~\bibnamefont{Toner}} \bibnamefont{and}
  \bibinfo{author}{\bibfnamefont{Y.}~\bibnamefont{Tu}}, \bibinfo{journal}{Phys.
  Rev. E} \textbf{\bibinfo{volume}{58}}, \bibinfo{pages}{4828}
  (\bibinfo{year}{1998}).

\bibitem[{\citenamefont{Toner et~al.}(2005)\citenamefont{Toner, Tu, and
  Ramaswamy}}]{Ramaswamy2005}
\bibinfo{author}{\bibfnamefont{J.}~\bibnamefont{Toner}},
  \bibinfo{author}{\bibfnamefont{Y.}~\bibnamefont{Tu}}, \bibnamefont{and}
  \bibinfo{author}{\bibfnamefont{S.}~\bibnamefont{Ramaswamy}},
  \bibinfo{journal}{Ann. Phys.} \textbf{\bibinfo{volume}{318}},
  \bibinfo{pages}{170} (\bibinfo{year}{2005}).

\bibitem[{\citenamefont{Ramaswamy et~al.}(2003)\citenamefont{Ramaswamy, Simha,
  and Toner}}]{Toner2003}
\bibinfo{author}{\bibfnamefont{S.}~\bibnamefont{Ramaswamy}},
  \bibinfo{author}{\bibfnamefont{R.~A.} \bibnamefont{Simha}}, \bibnamefont{and}
  \bibinfo{author}{\bibfnamefont{J.}~\bibnamefont{Toner}},
  \bibinfo{journal}{Europhys. Lett.} \textbf{\bibinfo{volume}{62}},
  \bibinfo{pages}{196} (\bibinfo{year}{2003}).

\bibitem[{\citenamefont{Chate et~al.}(2006)\citenamefont{Chate, Ginelli, and
  Montagne}}]{Montagne2006}
\bibinfo{author}{\bibfnamefont{H.}~\bibnamefont{Chate}},
  \bibinfo{author}{\bibfnamefont{F.}~\bibnamefont{Ginelli}}, \bibnamefont{and}
  \bibinfo{author}{\bibfnamefont{R.}~\bibnamefont{Montagne}},
  \bibinfo{journal}{Phys. Rev. Lett.} \textbf{\bibinfo{volume}{96}},
  \bibinfo{pages}{180602} (\bibinfo{year}{2006}).

\bibitem[{\citenamefont{Aranson and Tsimring}(2003)}]{Tsimring2003}
\bibinfo{author}{\bibfnamefont{I.}~\bibnamefont{Aranson}} \bibnamefont{and}
  \bibinfo{author}{\bibfnamefont{L.}~\bibnamefont{Tsimring}},
  \bibinfo{journal}{Phys. Rev. E} \textbf{\bibinfo{volume}{67}},
  \bibinfo{pages}{021305} (\bibinfo{year}{2003}).

\bibitem[{\citenamefont{Aranson et~al.}(2007)\citenamefont{Aranson, Volfson,
  and Tsimring}}]{Tsimring2007}
\bibinfo{author}{\bibfnamefont{I.~S.} \bibnamefont{Aranson}},
  \bibinfo{author}{\bibfnamefont{D.}~\bibnamefont{Volfson}}, \bibnamefont{and}
  \bibinfo{author}{\bibfnamefont{L.~S.} \bibnamefont{Tsimring}},
  \bibinfo{journal}{Phys. Rev. E} \textbf{\bibinfo{volume}{75}},
  \bibinfo{pages}{051301} (\bibinfo{year}{2007}).

\bibitem[{\citenamefont{Tu et~al.}(1998)\citenamefont{Tu, Toner, and
  Ulm}}]{Ulm1998}
\bibinfo{author}{\bibfnamefont{Y.}~\bibnamefont{Tu}},
  \bibinfo{author}{\bibfnamefont{J.}~\bibnamefont{Toner}}, \bibnamefont{and}
  \bibinfo{author}{\bibfnamefont{M.}~\bibnamefont{Ulm}},
  \bibinfo{journal}{Phys. Rev. Lett.} \textbf{\bibinfo{volume}{80}},
  \bibinfo{pages}{4819} (\bibinfo{year}{1998}).

\bibitem[{\citenamefont{Simha and Ramaswamy}(2002)}]{Ramaswamy2002}
\bibinfo{author}{\bibfnamefont{R.}~\bibnamefont{Simha}} \bibnamefont{and}
  \bibinfo{author}{\bibfnamefont{S.}~\bibnamefont{Ramaswamy}},
  \bibinfo{journal}{Phys. Rev. Lett.} \textbf{\bibinfo{volume}{89}},
  \bibinfo{pages}{058101} (\bibinfo{year}{2002}).

\bibitem[{\citenamefont{Baskaran and Marchetti}(2008)}]{Marchetti2007}
\bibinfo{author}{\bibfnamefont{A.}~\bibnamefont{Baskaran}} \bibnamefont{and}
  \bibinfo{author}{\bibfnamefont{M.~C.} \bibnamefont{Marchetti}},
  \bibinfo{journal}{Phys. Rev. E} \textbf{\bibinfo{volume}{77}},
  \bibinfo{pages}{011920} (\bibinfo{year}{2008}).

\bibitem[{\citenamefont{Blair et~al.}(2003)\citenamefont{Blair, Neicu, and
  Kudrolli}}]{Kudrolli2003}
\bibinfo{author}{\bibfnamefont{D.}~\bibnamefont{Blair}},
  \bibinfo{author}{\bibfnamefont{T.}~\bibnamefont{Neicu}}, \bibnamefont{and}
  \bibinfo{author}{\bibfnamefont{A.}~\bibnamefont{Kudrolli}},
  \bibinfo{journal}{Phys. Rev. E} \textbf{\bibinfo{volume}{67}},
  \bibinfo{pages}{031303} (\bibinfo{year}{2003}).

\bibitem[{\citenamefont{Galanis et~al.}(2006)\citenamefont{Galanis, Harries,
  Sackett, Losert, and Nossal}}]{Nossal2006}
\bibinfo{author}{\bibfnamefont{J.}~\bibnamefont{Galanis}},
  \bibinfo{author}{\bibfnamefont{D.}~\bibnamefont{Harries}},
  \bibinfo{author}{\bibfnamefont{D.}~\bibnamefont{Sackett}},
  \bibinfo{author}{\bibfnamefont{W.}~\bibnamefont{Losert}}, \bibnamefont{and}
  \bibinfo{author}{\bibfnamefont{R.}~\bibnamefont{Nossal}},
  \bibinfo{journal}{Phys. Rev. Lett.} \textbf{\bibinfo{volume}{96}},
  \bibinfo{pages}{028002} (\bibinfo{year}{2006}).

\bibitem[{\citenamefont{Narayan et~al.}(2006)\citenamefont{Narayan, Menon, and
  Ramaswamy}}]{Menon2006}
\bibinfo{author}{\bibfnamefont{V.}~\bibnamefont{Narayan}},
  \bibinfo{author}{\bibfnamefont{N.}~\bibnamefont{Menon}}, \bibnamefont{and}
  \bibinfo{author}{\bibfnamefont{S.}~\bibnamefont{Ramaswamy}},
  \bibinfo{journal}{J. Stat. Mech.}  (\bibinfo{year}{2006}).

\bibitem[{\citenamefont{Narayan et~al.}(2007)\citenamefont{Narayan, Ramaswamy,
  and Menon}}]{Menon2007}
\bibinfo{author}{\bibfnamefont{V.}~\bibnamefont{Narayan}},
  \bibinfo{author}{\bibfnamefont{S.}~\bibnamefont{Ramaswamy}},
  \bibnamefont{and} \bibinfo{author}{\bibfnamefont{N.}~\bibnamefont{Menon}},
  \bibinfo{journal}{Science} \textbf{\bibinfo{volume}{317}}
  (\bibinfo{year}{2007}).

\bibitem[{\citenamefont{Kudrolli et~al.}(2008)\citenamefont{Kudrolli, Lumay,
  Volfson, and Tsimring}}]{TsimringPolar}
\bibinfo{author}{\bibfnamefont{A.}~\bibnamefont{Kudrolli}},
  \bibinfo{author}{\bibfnamefont{G.}~\bibnamefont{Lumay}},
  \bibinfo{author}{\bibfnamefont{D.}~\bibnamefont{Volfson}}, \bibnamefont{and}
  \bibinfo{author}{\bibfnamefont{L.~S.} \bibnamefont{Tsimring}},
  \bibinfo{journal}{Phys. Rev. Lett.}  (\bibinfo{year}{2008}).

\bibitem[{\citenamefont{Aranson et~al.}(2008)\citenamefont{Aranson, Snezhko,
  Olafsen, and Urbach}}]{UrbachComment}
\bibinfo{author}{\bibfnamefont{I.}~\bibnamefont{Aranson}},
  \bibinfo{author}{\bibfnamefont{A.}~\bibnamefont{Snezhko}},
  \bibinfo{author}{\bibfnamefont{J.}~\bibnamefont{Olafsen}}, \bibnamefont{and}
  \bibinfo{author}{\bibfnamefont{J.}~\bibnamefont{Urbach}},
  \bibinfo{journal}{Science} \textbf{\bibinfo{volume}{320}},
  \bibinfo{pages}{612c} (\bibinfo{year}{2008}).

\bibitem[{\citenamefont{Narayan et~al.}(2008)\citenamefont{Narayan, Ramaswamy,
  and Menon}}]{MenonResponse}
\bibinfo{author}{\bibfnamefont{V.}~\bibnamefont{Narayan}},
  \bibinfo{author}{\bibfnamefont{S.}~\bibnamefont{Ramaswamy}},
  \bibnamefont{and} \bibinfo{author}{\bibfnamefont{N.}~\bibnamefont{Menon}},
  \bibinfo{journal}{Science} \textbf{\bibinfo{volume}{320}},
  \bibinfo{pages}{612d} (\bibinfo{year}{2008}).

\bibitem[{\citenamefont{Perrin}(1934)}]{Perrin1934}
\bibinfo{author}{\bibfnamefont{F.}~\bibnamefont{Perrin}}, \bibinfo{journal}{J.
  Phys. Radium} \textbf{\bibinfo{volume}{V}}, \bibinfo{pages}{497}
  (\bibinfo{year}{1934}).

\bibitem[{\citenamefont{Perrin}(1936)}]{Perrin1936}
\bibinfo{author}{\bibfnamefont{F.}~\bibnamefont{Perrin}}, \bibinfo{journal}{J.
  Phys. Radium} \textbf{\bibinfo{volume}{VII}}, \bibinfo{pages}{1}
  (\bibinfo{year}{1936}).

\bibitem[{\citenamefont{Han et~al.}(2006)\citenamefont{Han, Alsayed, Nobili,
  Zhang, Lubensky, and Yodh}}]{Yodh2006}
\bibinfo{author}{\bibfnamefont{Y.}~\bibnamefont{Han}},
  \bibinfo{author}{\bibfnamefont{A.~M.} \bibnamefont{Alsayed}},
  \bibinfo{author}{\bibfnamefont{M.}~\bibnamefont{Nobili}},
  \bibinfo{author}{\bibfnamefont{J.}~\bibnamefont{Zhang}},
  \bibinfo{author}{\bibfnamefont{T.~C.} \bibnamefont{Lubensky}},
  \bibnamefont{and} \bibinfo{author}{\bibfnamefont{A.~G.} \bibnamefont{Yodh}},
  \bibinfo{journal}{Science} \textbf{\bibinfo{volume}{314}},
  \bibinfo{pages}{626} (\bibinfo{year}{2006}).

\bibitem[{\citenamefont{Ojha et~al.}(2004)\citenamefont{Ojha, Lemieux, Dixon,
  Liu, and Durian}}]{Durian2004}
\bibinfo{author}{\bibfnamefont{R.~P.} \bibnamefont{Ojha}},
  \bibinfo{author}{\bibfnamefont{P.~A.} \bibnamefont{Lemieux}},
  \bibinfo{author}{\bibfnamefont{P.~K.} \bibnamefont{Dixon}},
  \bibinfo{author}{\bibfnamefont{A.~J.} \bibnamefont{Liu}}, \bibnamefont{and}
  \bibinfo{author}{\bibfnamefont{D.~J.} \bibnamefont{Durian}},
  \bibinfo{journal}{Nature} \textbf{\bibinfo{volume}{427}},
  \bibinfo{pages}{521} (\bibinfo{year}{2004}).

\bibitem[{\citenamefont{Ojha et~al.}(2005)\citenamefont{Ojha, Abate, and
  Durian}}]{Ojha2005}
\bibinfo{author}{\bibfnamefont{R.~P.} \bibnamefont{Ojha}},
  \bibinfo{author}{\bibfnamefont{A.~R.} \bibnamefont{Abate}}, \bibnamefont{and}
  \bibinfo{author}{\bibfnamefont{D.~J.} \bibnamefont{Durian}},
  \bibinfo{journal}{Phys. Rev. E} \textbf{\bibinfo{volume}{71}},
  \bibinfo{pages}{016313} (\bibinfo{year}{2005}).

\bibitem[{\citenamefont{Abate and Durian}(2005)}]{Abate2005}
\bibinfo{author}{\bibfnamefont{A.~R.} \bibnamefont{Abate}} \bibnamefont{and}
  \bibinfo{author}{\bibfnamefont{D.~J.} \bibnamefont{Durian}},
  \bibinfo{journal}{Phys. Rev. E} \textbf{\bibinfo{volume}{72}},
  \bibinfo{pages}{031305} (\bibinfo{year}{2005}).

\bibitem[{\citenamefont{Rajagopalan and Antonia}(2005)}]{Antonia}
\bibinfo{author}{\bibfnamefont{S.}~\bibnamefont{Rajagopalan}} \bibnamefont{and}
  \bibinfo{author}{\bibfnamefont{R.}~\bibnamefont{Antonia}},
  \bibinfo{journal}{Exp. Fluids} \textbf{\bibinfo{volume}{38}},
  \bibinfo{pages}{393} (\bibinfo{year}{2005}).

\bibitem[{\citenamefont{Wright et~al.}(2006)\citenamefont{Wright, Swift, and
  King}}]{King2006}
\bibinfo{author}{\bibfnamefont{H.}~\bibnamefont{Wright}},
  \bibinfo{author}{\bibfnamefont{M.}~\bibnamefont{Swift}}, \bibnamefont{and}
  \bibinfo{author}{\bibfnamefont{P.}~\bibnamefont{King}},
  \bibinfo{journal}{Phys. Rev. E} \textbf{\bibinfo{volume}{74}},
  \bibinfo{pages}{061309} (\bibinfo{year}{2006}).

\bibitem[{\citenamefont{Blair and Kudrolli}(2003)}]{Blair2003}
\bibinfo{author}{\bibfnamefont{D.}~\bibnamefont{Blair}} \bibnamefont{and}
  \bibinfo{author}{\bibfnamefont{A.}~\bibnamefont{Kudrolli}},
  \bibinfo{journal}{Phys. Rev. E} \textbf{\bibinfo{volume}{67}},
  \bibinfo{pages}{041301} (\bibinfo{year}{2003}).

\bibitem[{\citenamefont{Tsai et~al.}(2003)\citenamefont{Tsai, Ye, Rodriguez,
  Gollub, and Lubensky}}]{GollubChiral}
\bibinfo{author}{\bibfnamefont{J.}~\bibnamefont{Tsai}},
  \bibinfo{author}{\bibfnamefont{F.}~\bibnamefont{Ye}},
  \bibinfo{author}{\bibfnamefont{J.}~\bibnamefont{Rodriguez}},
  \bibinfo{author}{\bibfnamefont{J.}~\bibnamefont{Gollub}}, \bibnamefont{and}
  \bibinfo{author}{\bibfnamefont{T.}~\bibnamefont{Lubensky}},
  \bibinfo{journal}{Phys. Rev. Lett.} \textbf{\bibinfo{volume}{94}}
  (\bibinfo{year}{2003}).

\bibitem[{\citenamefont{Alder and Wainwright}(1967)}]{Alder1967}
\bibinfo{author}{\bibfnamefont{B.~F.} \bibnamefont{Alder}} \bibnamefont{and}
  \bibinfo{author}{\bibfnamefont{T.~E.} \bibnamefont{Wainwright}},
  \bibinfo{journal}{Phys. Rev. Lett.} \textbf{\bibinfo{volume}{18}},
  \bibinfo{pages}{988} (\bibinfo{year}{1967}).

\end{thebibliography}

\end{document}